An Investigation into Labouchère's Betting System to Improve Odds of Favorable Outcomes to Generate a Positive Externality Empirically

Jake Billings, Sebastian Del Barco

University of Colorado Denver

## ABSTRACT

The Labouchère gambling system is hypothesized to increase the probability of winning a predetermined arbitrary profit in a gambling system such as a coin flip or a roulette game in which both payouts and odds are 1:1. However, use of the system increases the downside monetary risk in the event of a streak of multiple losses. To begin, a player creates an arbitrary series of consecutive integers with a sum equal to the desired profit from multiple rounds of betting. Using the system, a player will either win an amount equal to the sum of the elements of the initial series or lose all of their available capital. This sequence was simulated multiple times to determine the statistical characteristics of both the return and of the loss in an average round of betting. By running the simulations of millions of rounds of Labouchere, it was possible to discern the probable outcomes of running the system using the Labouche gambling sequence and plotting the results on a graph to map the average return on the initial capital investment. The Labouchère system is very psychologically appealing to players because when applied over time it provides very consistent linear returns. However, there is eventually a critical moment at which the available capital for betting is exceeded and a player loses all of their available capital. It was found that as the number of bets increased, the outcome of applying the sequence approached zero.

# INTRODUCTION

The Labouchère system is designed for zero sum betting systems in which the chance of winning is 50%. This can be thought of as two players betting equal amounts of money on the outcome of a coin flip. The winner of the bet receives the money he or she put down as well as the money the other player put down. It is known, empirically, that the probability a fair coin toss will return an approximately equal amount of heads and tails as a coin toss approaches infinity, (Downton, F., & Hill, J. M. 1982). The Labouchère betting sequence strategy attempts to increase the chances of a positive return over multiple rounds of betting. This is then applied to a wagering system, such as coin flip, sports outcome, or online casino. Using this strategy, the player decides the desired return ahead of time by writing a sequence of consecutive integers where the sum is the desired return from multiple rounds of betting. Once this initial sequence is selected, the player begins betting. The system can be implemented as a recursive algorithm. Recursion terminates in either the event that the sequence is empty or the event that the player runs out of available capital. On each recursion, a bet is made of value equal to the sum of the first and last numbers of the sequence. If the length of the sequence is one, then the bet is equal to the sole member of the sequence. If the bet is won, then the first and last members are spliced from the sequence and the next round begins. However, if the bet results in a loss, then an integer equal the size of the lost bet is appended to the sequence and the next round begins. As determined by the parameters for termination of recursion, the only cases in which the algorithm will terminate are those in which the player has either won an amount equal to the summation of the original sequence or has lost all of their available capital.

The Labouchere system is thought to return a profit in cases such that a player wins more bets than they lose. (Downton, F. 1980). This logic is flawed as it perpetuates The Gambling Fallacy. (Gray, H., LaPlante, D., & Shaffer, H. 2012). It is not possible to improve betting outcomes through the use of a betting strategy. It is possible to profit from the use of this betting system. The system works by increasing bet size when farther from the desired return. As a result, the risk of a loss increases with the number of bets performed. (Savaş, E., & Patterson, R. F. 2007). The gambling strategy increases the stakes with every loss that is experienced. (Schmidt, K., Winterhof, A., & SpringerLink 2014). This results in a greater chance that the size of a bet will exceed a player's available capital as time passes. In other words, the probability of losing increases as more rounds are played. As the number of rounds played approaches infinity, the probability of a loss approaches one. Players implementing this strategy will continue until they have insufficient funds and therefore cannot play the game. Players experience gambler's fallacy when implementing this strategy because this sequence strategy results in relatively few losses relative to the number of wins. However, this strategy amplifies the magnitude of the loss experienced which outweighs the moderate, although relatively consistent, wins that are experienced. (Gray, H., LaPlante, D., & Shaffer, H. 2012). This is indicative of the negative progressive system of the Labouchere sequence which facilitates losses and perpetuates the gambler's fallacy. (Makri, F. S., & Psillakis, Z. M. 2016).

The algorithm continues until one of the following criteria for termination are met: the player can no longer bet due to insufficient funds or the sequence is empty. At the point of algorithmic termination, one of two outcomes has occurred. Either the player has lost all available capital or the player has profited by an amount equal to the summation of the original starting sequence. This bet must be reached unless the probability of a loss exceeds 2/3 or where $\rho = .66$ (Downton, F., & Hill, J. M. 1982). The fatal flaw of this betting system is that the loss will eventually exceed the gambler's available capital and thus will result in a complete loss. The purpose of this investigation was to determine the probabilistic outcome by testing empirical results via simulation of prolonged use of the Labouchère betting system.

## METHODS

Implementation of Random Coin Flips in Python

For the purpose of simulating random events in Python, getrandbits() function the random module was used. According to python documentation, the underlying entropy is extracted from the Mersenne Twister algorithm written in C, (Python Standard Library). It is worth noting that the random events used for simulation of millions of bets were acquired from a pseudo-random number generator. This implementation balanced computational efficiency, speed, and practicality.

```python
import random

# Return a random boolean in order to simulate a coin flip.
# This is approximately 50% odds.
# The odds are not 50% on online betting sites. They typically take a "house advantage," so the odds are actually
# similar to 49.95%.
def flip_coin():
    return bool(random.getrandbits(1))
```

Recursive Implementation of the Labouchere betting system in Python

```python
# Runs a simulation of the Labouchere betting system with a given starting sequence and
balance.
# Returns the ending balance after running the system to completion and the number of bets it
took
# This is a recursive function. Each function call is one "round" of betting.
#
# See: https://en.wikipedia.org/wiki/Labouch%C3%A8re_system
def gamble(sequence, balance):
    # If the sequence is empty, the labouchere system says that the round is over.
    # End the recursion. This is essentially a win.
    if len(sequence) < 1:
        # Return 0 and the initial balance because no bet was made
        return 0, balance

    # If the sequence is of length 1, the bet is the number in the sequence. Otherwise, it is the
first number
    # added to the last number.
    if len(sequence) is 1:
        bet = sequence[0]
    else:
        bet = sequence[0] + sequence[-1]

    # You can't bet more money than you have (this isn't Wall Street), so
    # betting more than the initial balance ends recursion. This is essentially a loss.
    if bet > balance:
        # Return 0 and the initial balance because no bet was made
        return 0, balance

    # If a random boolean is true, we won.
    won = flip_coin()

    # Add or subtract from the balance based on the result of the bet and then play the next
round.
    # Labouchere states that the first and last numbers of the sequence are removed in the event
of a win, and the
    # amount of the bet is added to the end of the sequence in the event of a loss.
    if won:
        bets, resulting_balance = gamble(sequence[1:-1], balance+bet)
        # Increment the number of bets because we made a one and return the resulting balance
        return bets+1, resulting_balance
    else:
        bets, resulting_balance = gamble(sequence+[bet], balance-bet)
```

```python
# Increment the number of bets because we made a one and return the resulting balance
return bets+1, resulting_balance
```

The following are examples of the round sequences that were simulated 10,000 and 10,000,000 times:

Example Round 1: Best Possible Case

Initial Round:  {$1, $2, $3}
Bet 1:  $1 + $3 = $4
*Assume Bet 1 is won.*
Total Profit: $4

Sequence after Bet 1: {$2}
Bet 2: $2
*Assume Bet 2 is won.*
Total Profit: $6

Sequence after Bet 2: ∅
*Algorithm completes in 2 bets.*

Example Round 2: Worst Possible Case

Initial Round:  { $1, $2, $3 }
Bet 1:  $1 + $3 = $4
*Assume Bet 1 is lost.*
Total Loss: $4

Sequence after Bet 1: { $1, $2, $3, $4 }
Bet 2: $1 + $4 = $5
*Assume Bet 2 is lost.*
Total Loss: $9

Sequence after Bet 2: { $1, $2, $3, $4, $5 }
Bet 3: $1 + $5 = $6
*Assume Bet 3 is lost.*
Total Loss: $15

Sequence after Bet 3: { $1, $2, $3, $4, $5, $6 }
Bet 4: $1 + $6 = $7
*Assume Bet 4 is lost.*

Total Loss: $22

…

Sequence after Bet ∞: { <u>$1</u>, $2, $3, $4, $5, $6 … <u>$∞</u> }

Bet ∞: $∞

Total Loss: $∞

*Algorithm does not complete if no bets are won.*

The algorithm will always end return a profit of $6 if there is no limit to the size of the bet. The algorithm never completes until the profit is made.  The profit is always equal to the sum of the original sequence.

$1 + $2 + $3 = $6

In a practical setting, the algorithm will often halt because it exceeds the available funds of the player and is unable to make a bet. As a result, wealthier players are more likely to make a $6 profit when running the algorithm. However, players risk losing the complete contents of their bank account since that is the only other case in white the algorithm will complete.

Implementation of the Bankroll Strategy in Python

The bankroll strategy, also known as the Zimmerman Strategy, takes the aggregate of multiple rounds of Labouchere betting. Supposedly by making bets much smaller than one's bankroll, one can maximize the probability of winning because it is very unlikely that Labouchere will result in the loss of the entire bankroll all at once.

```python
from labouchere import gamble

# After a round of Labouchere, if a balance is above a certain threshold, all money above that threshold is removed
# from the bankroll as profit, and it is never used for gambling again.
#
# sequence The initial sequence to use in every round of Labouchere
# rounds The number of rounds of Labouchere to run
# max_balance The threshold above which profits are extracted from the bankroll (balance)
# initial_balance The initial size of bankroll (balance) to use
def run_bankroll_strategy(sequence=[1,2,3], rounds=5, max_balance=6000,
initial_balance=4000):
    # Calculate the initial bet that Labouchere will make
    initial_bet = sequence[0]
    if len(sequence) > 1:
        initial_bet += sequence[-1]

    # Store the current balance in a variable
    balance = initial_balance

    # Store the "extracted profit" or "money scraped off the top"
    extracted_profit = 0

    # Run the number of rounds stored in the variable rounds
    for i in range(1, rounds):
        # Stop playing if you're out of money.
        if balance<initial_bet:
            # print "You're broke."
            break

        # Run one full round of Labouchere
        bets, resulting_balance = gamble(sequence, balance)

        # Store the new balance from Labouchere
        balance = resulting_balance
```

```python
    # Scrape off the top in accordance with the bankroll strategy
    unwanted_money = balance-max_balance
    if unwanted_money > 0:
        balance -= unwanted_money
        extracted_profit += unwanted_money

    return (balance+extracted_profit)-initial_balance
```

Downsampling of logarithmic data

This listing is included to exemplify retrieval of experimental values for research. wins_to_bankroll_downsampled() monitors the time it takes for the algorithm complete as compared to the maximum time it could take complete. Additionally, it down samples data proportionally to the order of magnitude of the bankroll. This makes processing simulations with many large bankrolls far more efficient and far more manageable in Excel.

```python
# Analyzes the number of bets won
# returns a histogram where the number of rounds resulting in each number of bets is counted
# Down samples proportionally the order of magnitude of the bankroll
def wins_to_bankroll_downsampled(
        sequence=[1, 2, 3],
        min_bankroll=0,
        max_bankroll=400000,
        rounds_per_bankroll=10000,
        update_frequency=2,
        downsample_constant=10):
    # Store results in a dict
    results = [['balance', 'wins', 'losses', 'draws']]

    # Initialize variables for performance benchmarking
    start = time()
    last_update = 0

    total_rounds = (max_bankroll - min_bankroll) * rounds_per_bankroll

    step = 1
    balance = min_bankroll

    while balance < max_bankroll:
        wins = 0
        losses = 0
        draws = 0

        for i in range(0, rounds_per_bankroll):
            bets, resulting_balance = gamble(sequence, balance)

            if resulting_balance > balance:
                wins += 1
            elif balance > resulting_balance:
                losses += 1
            else:
```

```python
        draws += 1

        # Update a user on the progress of the simulation
        t = time()
        if t - last_update > update_frequency:
            print "Completed %s/<%s rounds in %s seconds. Step size: %s" % (
            balance / step * rounds_per_bankroll, total_rounds, floor(t - start), step)
            last_update = t

    results.append([balance, wins, losses, draws])

    balance += step
    step = floor(balance / downsample_constant)

    if step < 1:
        step = 1

    # Print a benchmark for how long the simulation took
    end = time()
    print "Done in %s seconds; Avg. %s seconds/round" % ((end - start), (end - start) /
    total_rounds)

    return results
```

Data Export and Analysis

Relevant data from simulations written as variants of the above gamble() and
run_bankroll_strategy() functions was exported to .csv files for analysis in Microsoft Excel.

```python
# Exports each key value pair as a row in a csv file
def export_dict_as_csv(d, name='export.csv'):
    with open(name, 'w') as wfile:
        for key in d:
            wfile.write(str(key)+','+str(d[key])+'\n')

# Exports a 2D array as a csv file
def export_array_as_csv(d, name='export.csv'):
    with open(name, 'w') as wfile:
        for row in d:
            for cell in row:
                wfile.write(str(cell)+',')
            wfile.write('\n')
```

Data was exported in csv files formatted similarly to the following:

```
balance, wins, losses, draws,
0, 0, 0, 1000,
1, 0, 0, 1000,
2, 0, 0, 1000,
3, 0, 0, 1000,
4, 371, 629, 0,
....
22, 739, 261, 0,
23, 754, 246, 0,
24, 760, 240, 0,
25, 775, 225, 0,
....
39996, 1000, 0, 0,
39997, 1000, 0, 0,
39998, 1000, 0, 0,
39999, 1000, 0, 0,
```

All data and code used in this research is available under the MIT license at
https://github.com/jake-billings/research-labouchere.

To assess the variance between the $R^2$ values in throughout the distribution of the length of sequences before the Labouchere sequence completes, it will be necessary to assess the homogeneity of a number of bets that were placed. To do this, it will be necessary to run a bivariate case of the Gaussian multivariate distribution to assess the correlated real value of the random variables. This case found the probability density which was used to apply a Wald Test for homogeneity of the distribution of the number of bets before the sequence was complete.

The Bivariate case was run using the following equation:

$$f(x, y) \ = \ \frac{1}{2\pi \cdot \sigma_X \cdot \sigma_Y \sqrt{1-\rho^2}} \cdot \left( -\frac{1}{2(1-\rho^2)} \left[ \frac{(x-\mu_x)^2}{\sigma_X^2} + \frac{(y-\mu_y)^2}{\sigma^2} - \frac{2\rho(x-\mu_X)(y-\mu_Y)}{\sigma_X \sigma_Y} \right] \right)$$

This was found with the following assumptions:

$$P = \text{Correlation between } X \text{ and } Y$$

$$P = (x_1, x_2) = \frac{V_I}{\sigma_1 \sigma_2}$$

$$\sigma_X > 0$$

$$\sigma_Y > 0$$

$$\boldsymbol{\mu} = \begin{pmatrix} \mu_X \\ \mu_Y \end{pmatrix}$$

$$\boldsymbol{\Sigma} = \begin{pmatrix} \sigma_X^2 & \rho \sigma_X \sigma_Y \\ \rho \sigma_X \sigma_Y & \sigma_Y^2 \end{pmatrix}$$

This determined the parameters of the correlation. After this was done it was possible to run the Wald test for homogeneity on the number of bets before the sequence was complete.

To run a Wald test, it was necessary to find the maximum likelihood test statistic for the distribution of the number of bets before completion. This was found using the Poisson distribution.

$$f(x_1, x_2, x_3, \ldots, x_n \mid \lambda) = \frac{e^{-\lambda}\lambda^{x_1}}{x_1!} \frac{e^{-\lambda}\lambda^{x_2}}{x_2!} \frac{e^{-\lambda}\lambda^{x_3}}{x_3!} \ldots \frac{e^{-\lambda}\lambda^{x_1}}{x_1!} = \frac{e^{-n\lambda}\lambda^{\sum x_i}}{x_1! x_2! x_3! \ldots x_n!}$$

The linearity of the logit as described by the logarithmic test statistics of the f distribution was found by using the following assumption test:

$$ln(f) = -n\lambda + (ln(\lambda) \sum x_i - ln(\prod x_i!)$$

This was supported by the assumption that the derivative with respect to the maximum likelihood statistical was zero.

$$\frac{dlnf}{d\lambda} = \frac{\sum x_i}{\lambda} - n = 0$$

Finally, the last statistic will hold true when the previous assumptions were found.

$$\widehat{\lambda} = \frac{\sum x_i}{n}$$

Once these assumptions were tested to find the maximum likelihood test statistic, it was possible to use the Wald test for homogeneity:

$$X^2 = (R\widehat{\theta}_n - r)'(R\widehat{\theta}_n - r)\left[R(\frac{\widehat{V}_n}{n})R'\right]^{-1}$$

Where:

$\widehat{V}_n$ = Covariance matrix estimator

$\widehat{\theta}_n$ = Sample Estimator of parameters

**RESULTS**

**Table 1**. Distribution of Bets Before Completion for 10,000 Rounds

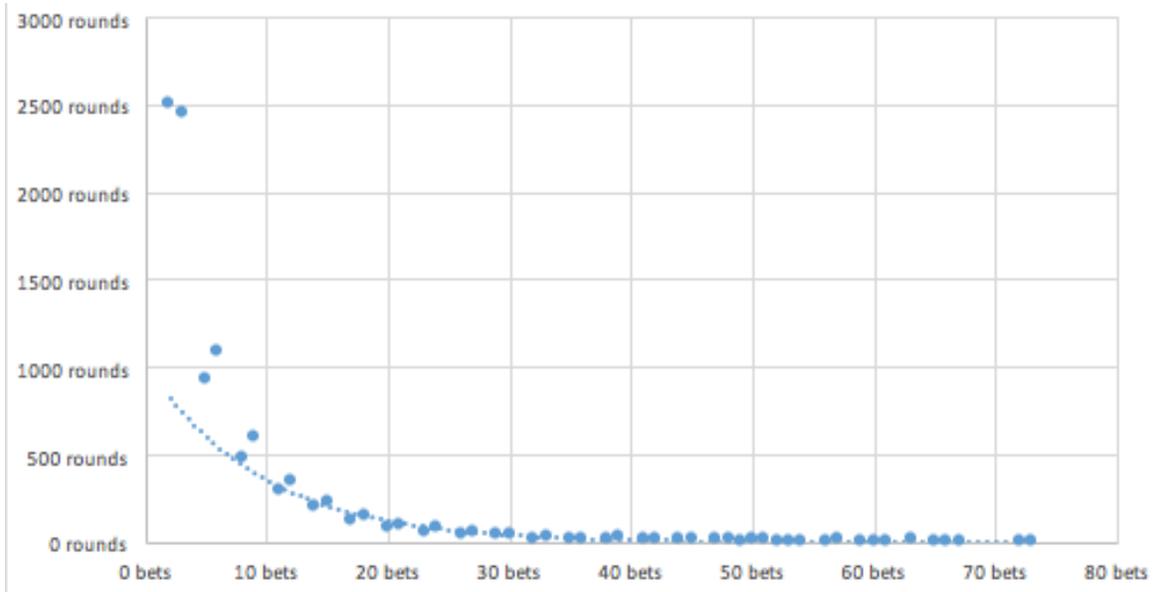

Table number 1 depicts a number of bets that were placed before the Labouchere sequence was terminated for 10,000 rounds. This was mapped with an exponential function y= $1026.3e^{-0.806x}$. This regression equation was found to account for 93.1% of the outcomes in this experiment which can be seen by the R-squared value of $R^2$= 0.93172. After a computer simulation completed in 15.821 seconds; Avg. 1.582e-06 seconds/round with an initial sequence of [$1,$2,$3] and an initial balance, or bank roll, of $4,000. Table number 2 below depicts the sample distribution with 10,000,000 rounds. The next table demonstrates the distribution of the bets placed before each round was completed.

**Table 2**. Distribution of Bets Before Completion for 10,000,000 Rounds

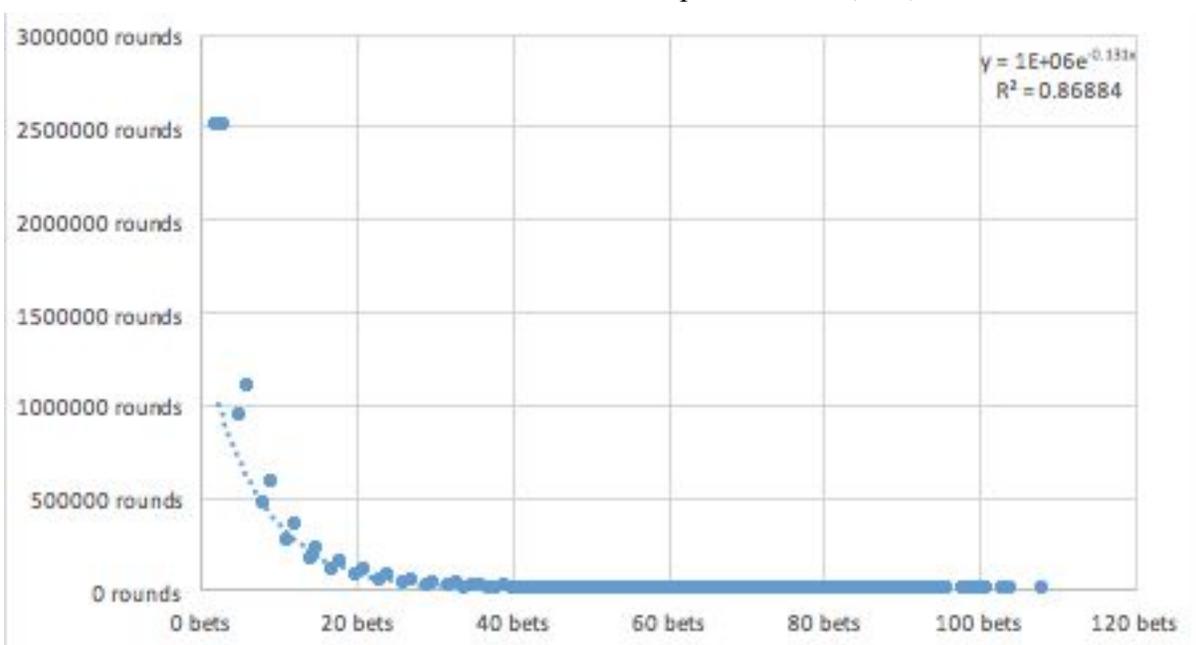

Table number 2 depicts a number of bets that were placed before the Labouchere sequence was terminated for 100,000,000 rounds. This was mapped with an exponential function y= $10^6 e^{-0.131x}$. This regression equation was found to account for 86.9% of the outcomes in this experiment which can be seen by the R-squared value of $R^2 = 0$. After a computer simulation completed in 85.821 seconds; Avg. 8.582e-06 seconds/round with an initial sequence of [$1, $2, $3] and an initial balance, or bank roll, of $4,000. The next table shows the games won relative to the size of the bankroll.

Based on a simulation of 10,000,000 rounds Labouchere betting, the following exponential decay regression can be used to model the number of rounds out of 10,000,000 that completed in a given number of bets with an $R^2$ value of 0.86884.

$$f(x) = 10^6 \cdot e^{-0.131x}$$

Where f(x) = is the number of rounds in 10,000,000 that can be expected to complete in x number of bets

To determine the probability that a single round will not complete within a given number of bets, divide the original function by 10,000,000. For instance, consider the datum at 2 bets. Approximately 2,500,000 rounds completed within approximately 2 bets.

$$g(x) = 0.1 \cdot e^{-0.131x}$$

Where g(x) = is the probability that a single round will not complete within a given number of bets in x number of bets

This is consistent with the shortest/best-case completion of an initial sequence of [1, 2, 3] in which a bet of $4 and then a bet of $2 are both won considering theoretical probability and regressed predictions from experimental results. $2.5 \cdot 10^6$ out of $1.0 \cdot 10^7$ rounds, or 25% of [1, 2, 3] rounds should complete within 2 bets. $f(2) \approx 2.5 \cdot 10^6$ ($\rho = \pm .7$). Including numerical regression inaccuracy, $f(2) = 7.69 \cdot 10^5$. $g(2) \approx .25$ ($\rho = \pm .7$). Including numerical regression inaccuracy, $g(2) = .076$.

As a result, the function $g(x) = 0.1 \cdot e^{-0.131x}$ can be used to model the probability of the termination of the recursive Labouchere betting algorithm with a given number of bets.

Furthermore, the probability, $\rho$, of that the algorithm has terminated after a given number of bets can be computed by integrating g.

$$\rho(x) = \int_0^x g(x) + 1$$

$$\rho(x) = \int_0^x 0.1 \cdot e^{-0.131x} + 1$$

$$\rho(x) = -0.131 \cdot e^{-0.131x} + 1$$

It is expected that as the number of bets approaches infinity, the probability of algorithmic termination approaches 1.

$$\lim_{x \to \infty} \rho(x) = 1$$

With numeric regression inaccuracy of $\rho = \pm .7$,

$$\lim_{x \to \infty} \rho(x) \approx 0.763$$

**Table 3**. Games Won vs. Bank Roll Size

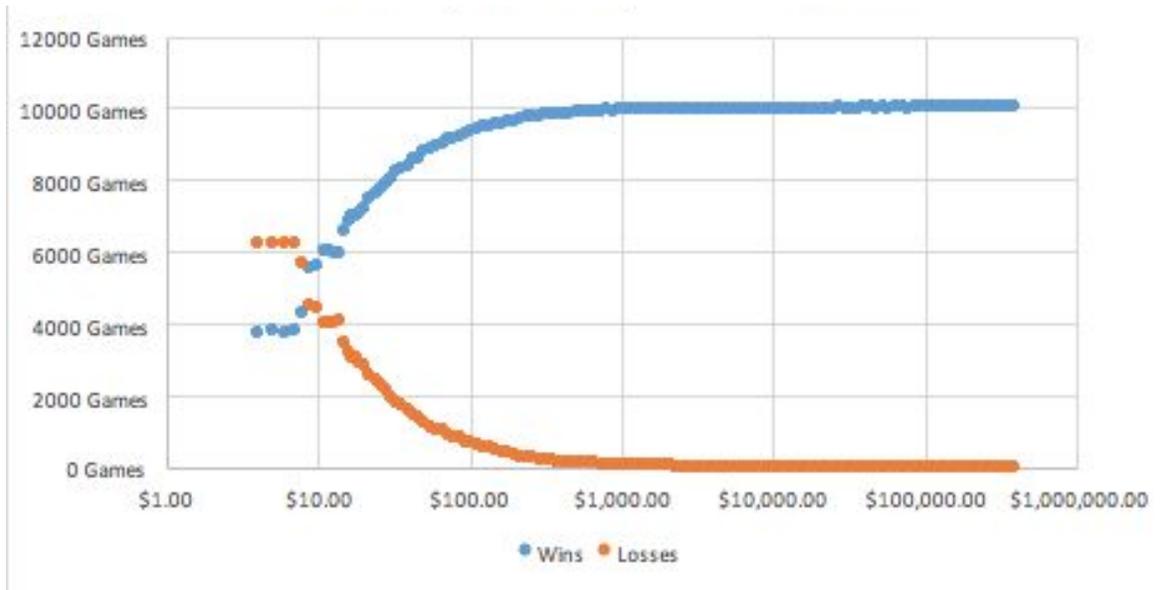

Table number 3 depicts the amount of games that were won vs. the amount of games that were lost relative to each other with respect to the size of the bank roll or the initial capital. This was done for multiple tests with 1000 games or sequences run. This was done for a bankroll size that started with a bankroll of $5 to a bankroll of $500,000. The next table depicts the capital balance relative to the amount of bets taken.

Based on analysis of the probability tree, it is clear that players with infinite available capital should always profit from the Labouchère betting system. All branches eventually lead to a profit of the sum of the initial sequence if the player never runs out of capital. Consider a function that models the probability of profiting from a round of Labouchère. The limit of ρ(x) as x approaches infinity should be 1.

$$\lim_{x \to \$\infty} \rho(x) = 1$$

Additionally, ρ(x) should be modeled by the experimental data collected in the simulation that includes 10,000 bets at each bankroll value between $0 and $1.0 · 10^6. Consider $f$(x), which is the number of simulated rounds won out of 10,000 rounds with an initial sequence of [$1, $2, $3] and available capital $x$.

$$f(x) \approx 10,000 \cdot \rho(x)$$

Ergo,

$$\lim_{x \to \$\infty} f(x) = 10,000$$

As a result, regarding the probability of success in relation to available capital, the experimental results are consistent with theoretical predictions. Larger amounts of available capital increase the probability of a positive return using the system.

**Table 4**. Capital Balance vs. Number of Bets Placed

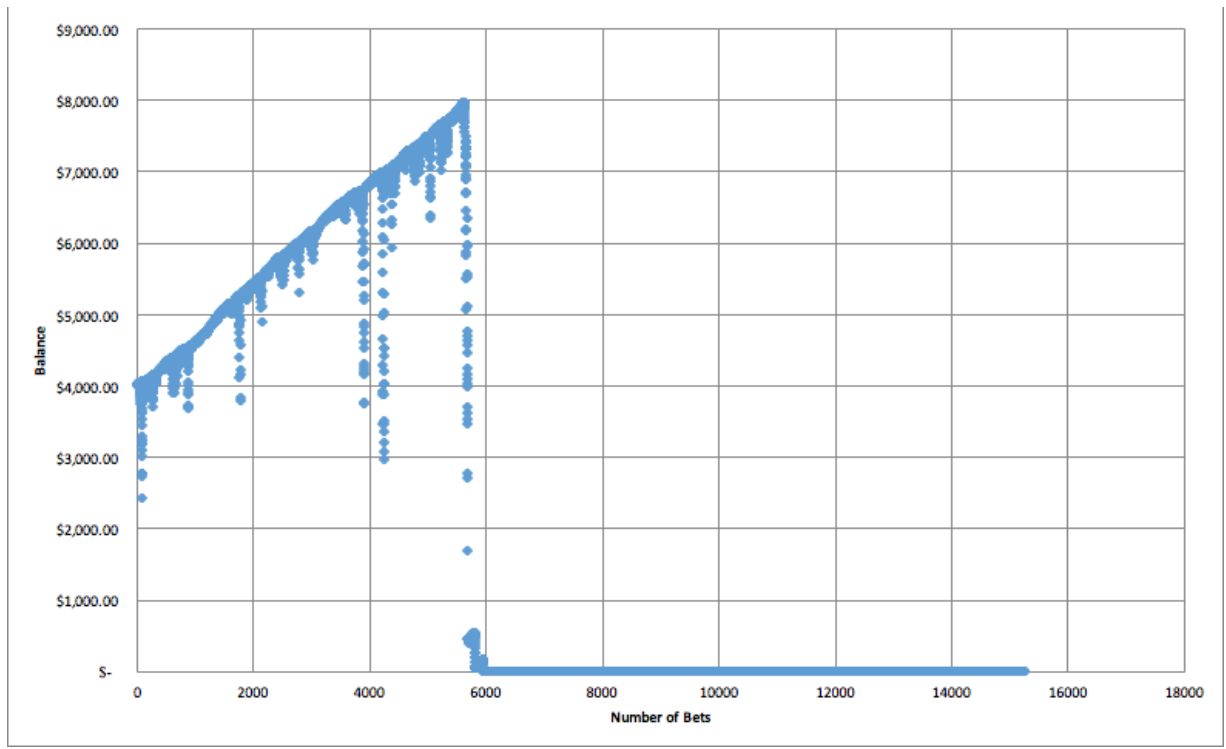

Table number 4 depicts the amount of capital available to a player in simulated dollars vs. the number of bets that were placed for one simulation of repeated Labouchère. Every time a round terminated, another began with the new balance. This was shown for bets that were made with an initial capital of $4,000. It is apparent from the graph, that a player received almost linear increase in capital until almost 6,000 bets were placed. At this point, the player faced too many sequential losses. As a result, the bet size required to continue exceeded available capital, and the player lost all available funds. This graph exemplifies the psychological temptation players may feel to continue playing given that their profit appears very consistent despite occasional dips. the next table demonstrates a linear profit that is generated from the bankroll strategy.

Consider a function $b(x)$ that describes the balance of a player's bankroll after x bets for the above simulation.

$$x \in \mathbb{Z}$$

$$b(0) = b_i = \$4,000$$

The value of $b(x)$ at any x depends on the probabilistic behavior of the bets placed. However,

$$\lim_{x \to \infty} b(x) = \$0$$

Because eventually, a loss streak will result in a bet size that exceeds a player's available capital. The bankroll strategy seeks to mitigate the risks of losing the entire bankroll by extracting profit once a player's bankroll, or available capital reaches a certain predetermined level.

**Table 5**. Profit Extracted from Bankroll Strategy

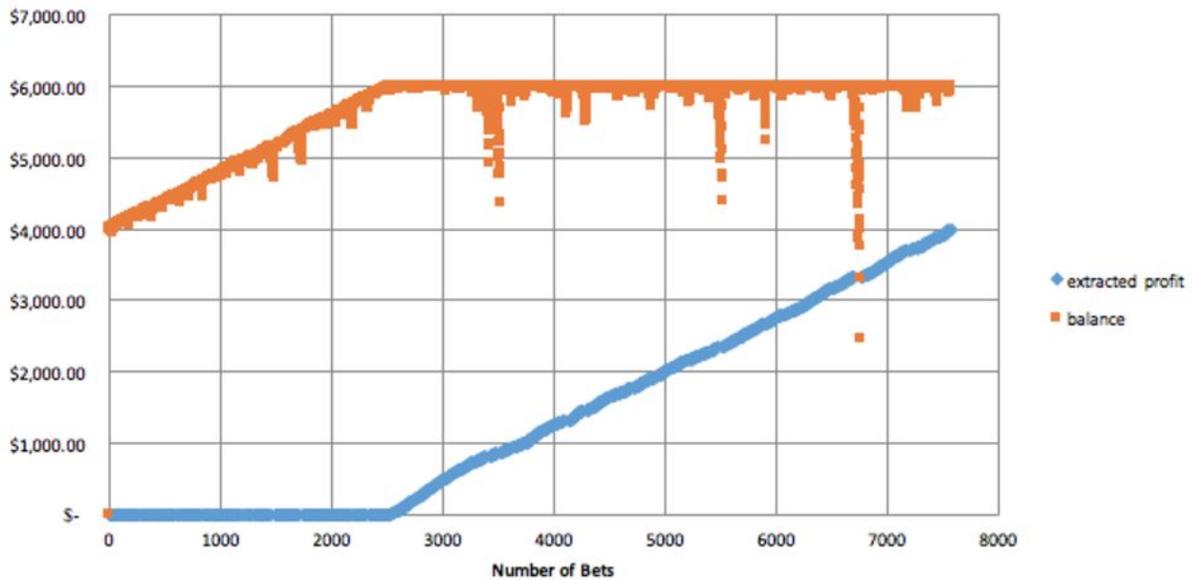

Table number 5 demonstrates the potential profitability of the bankroll strategy under favorable conditions of repeated Labouchére with a starting balance of $4,000 and a bankroll threshold of $6,000. Any profit the player made that brought the balance of the bankroll above $6,000 was extracted. These extracted funds grew steadily since they were no longer available for gambling. Profit for the bankroll strategy is calculated as the sum of the bankroll and the extracted profit. The next table describes the results of continuing the bankroll strategy and the loss that is eventually experienced.

$$p(x) = \text{extracted profit}$$

$b(x)$ = value of bankroll

$b(0)$ = the initial value of the bankroll

$P(x)$ = net profit

$P(x) = p(x) + b(x) - b(0)$

Where

$x$ = the number of bets performed

In this use of the strategy, the player profited approximately $6,000 in the 7500 bets the simulation ran for. However, the bankroll strategy does not guarantee positive returns.
Once the player's extracted profit is greater than the initial value of the bankroll, the player has profited and will not experience a net loss compared to their initial bankroll from further playing.

When $p(x) > b(0)$,

$P(x) > 0$.

**Table 6**. Loss Experienced with Bankroll Strategy

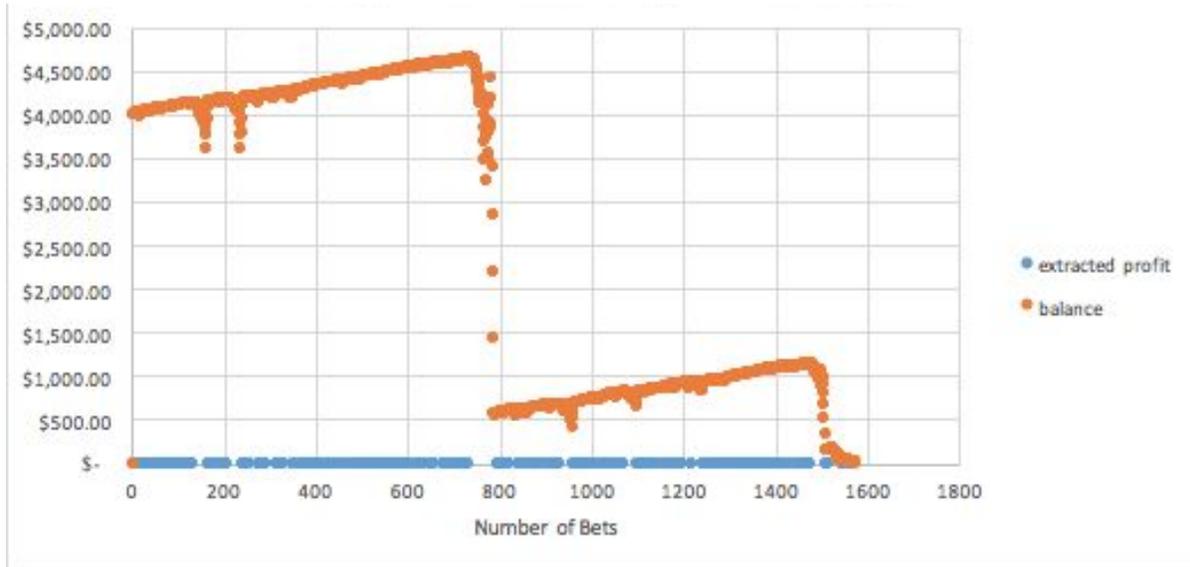

Table number 6 exemplifies a complete loss within 1600 bets despite perfect implementation of the bankroll strategy with a starting balance of $4,000 and a threshold of $6,000. The player experiences a losing streak near 700 bets before even reaching the threshold to extract profit from the bankroll. The initial loss streak nearly exceeds the player's bankroll. The player's balance then climbs from $500 to nearly $1,200 before experiencing another loss streak that brings the player's bankroll to 0. The next table demonstrates the results gathered from the test of homogeneity.

In other words, a player must double their money in order to guarantee a return from the bankroll strategy. However, if a player can double their money, their net profit is then guaranteed as the extracted profit is no longer used for gambling. However, it should be noted that players always risk the loss of their entire bankroll while gambling using the Labouchére system.

**Table 7**. Wald Test for Homogeneity

| Test Statistic | Value | Test | Significance |
|---|---|---|---|
| First R Square | 0.93172 | Homogeneity | .000 |
| Second R Square | 0.86884 | Homogeneity | .000 |

Table number 7 showcases the outcome of the test for homogeneity. This test was run to determine the difference and homogeneity of the first and second R-Squared value. Based on this test, with a confidence of 95% or at an alpha level of $\alpha = 0.05$, it was found that there was a statistically significant difference between the R-Squared value of the 10,000 sample simulation and the 10,000,000 sample simulation.

# DISCUSSION

This study determined used a computational simulation to predict the results of applying a Labouchere Sequence on a betting system with a dichotomous outcome of a gain or a loss. The simulation was run using Python which may have introduced confounding variables. The results that were produced were empirical results that were calculated by a pseudorandom generator which produced the results displayed. (Makri, F. S., & Psillakis, Z. M. 2016). . This may generate an inaccuracy in the results as the sample size for a number of bets that were gathered increased. This may have been quantified because of the change in accuracy of the $R^2$ value for the bankroll size of 10,000 bets to 10,000,000 bets. This change in $R^2$ resulted in a change in the accuracy of the measurement of about 6% which was directly affected by the increase in a number of rounds that were simulated. Overall, the inaccuracy of the computer simulation limits the application of results from this study because the inaccuracy is not quantified. Correspondingly, a future investigation would be required to specifically and accurately quantify the inaccuracy that is generated from the Labouchere sequence simulation that was created for this study.

The ability to quantify and therefore assess the relationship between the distribution of bets before the completion of a sequence is limited based on the observed decrease in accuracy. There was a significant different between the accuracy of $R^2$ value that was measured in both tests which were a different of 3 orders of magnitude. This was tested with the Wald test statistics for homogeneity which determined that the difference between the R-square value of the first and second distribution of bets. This confirmed the assumption that there the difference in accuracy changes with a significant change in sample size that was measured. Therefore, as the sample size increased the inaccuracy of the model decreased. Correspondingly, the variance explanation as found by the $R^2$ value decreases as the sample size increases.

# CONCLUSION

It was found that the algorithm for Labouchére completes nondeterministically. The probability of termination of a round can be modeled with the following exponential decay regression equation: $f(x) = 1026.3e^{-0.806x}$. However, it was found that the accuracy of this regression equation decreased by a statistically significant difference as tested by the homogeneity Wald Test when the sample size increased which limits its application. The probability that a Labouchere sequence will finish provided a number of bets was modeled again with a similar equation for the Labouchere sequence simulation for the test with 10,000,000 samples. The equation that was found using this sample size was found with the exponential decay function $f(x) = 10^6 e^{-0.131x}$. Although there was a statistically significant difference between the R-Squared value and therefore the accuracy, the conclusions drawn from the first test with 10,000 samples is consistent with the conclusion that can be drawn from the 10,000,000 sample simulation.

The Labouchere sequence strategy was simulated to determine whether or not a player will be able to generate a return on the gamble. It was found that only a player with an infinite amount of capital is capable of profiting from the Labouchere betting system. This was found because a player with an infinite amount of capital would be capable of surmounting the eventful loss pattern that was found by running the simulation. This loss pattern was averted in many cases where the sequence was terminated, however, it was determined a great enough loss pattern would result in the complete loss of the initial capital. A player with enough capital to surpass these loss patterns would then be capable of consistently generating a profit from the gamble. Therefore, this player would be capable of generating a profit equal to the sum of the initial sequence that is run. However, in a practical setting, a given player with some amount of capital would not be capable of generating a consistent profit equal to the sum of the first sequence in the long run because the player does not have enough capital. This was confirmed through the analysis of the bankroll size vs. a number of games that were won. It was shown that as the size of the bankroll increased, a number of games that were won correspondingly increased. This supported the practical conclusion drawn from this simulation that a player would not be able to circumvent the loss experienced from a game.

The gambler's fallacy that is experienced by many players was examined through this simulation through the analysis of bankroll and a number of bets that were placed in a game. It is known that a player in a game with any probability of losing, will eventually experience a loss. (Marmurek, H. H. C., Switzer, J., & D'Alvise, J. (2015). The structure of a Labouchere sequence magnifies the loss that players may experience. (Nuida, K., Abe, T., Kaji, S., Maeno, T., & Numata, Y. (2012). However, players experiencing gambler's fallacy may ignore this eventual reality which results in a loss of capital. This was investigated and confirmed to apply for the Labouchere sequence strategy because as a player continued to place bets, the loss experienced continue to increase and eventually terminated the amount of capital a player had. This was found when the graph that charted the amount of capital a player had vs. a number of bets that the player implemented confirmed the loss prediction. This was seen because the player was consistently making a profit in the sequence with observed losses, however, it was found that as the player continued this sequence, the player eventually faced a loss that resulted in a complete loss. Through the simulation, it was found that the Labouchere sequence strategy does not allow a player to avert a loss.

# CITATIONS